\documentclass[iop,apj,tighten]{emulateapj}
\usepackage{amsmath}
\usepackage{multirow}
\usepackage{colortbl}
\usepackage{float}
\usepackage{color}
\usepackage{graphicx}
\usepackage{hyperref}
\usepackage[numbered=true]{bookmark}
\usepackage{threeparttable}
\usepackage{subfigure}

\begin{document}

\shorttitle{Kepler AGN Variability Asymmetry}
\shortauthors{Chen \& Wang}

\title{Analyses on the Variability Asymmetry of Kepler AGNs}
\author{Xiao-Yang Chen \& Jun-Xian Wang}

\affil{CAS Key Laboratory for Research in Galaxies and Cosmology, Department of Astronomy, University of Science and Technology of China, Hefei, Anhui 230026, China; cosmos@mail.ustc.edu.cn, jxw@ustc.edu.cn}

\begin{abstract}
	The high quality light curves of Kepler space telescope make it possible to analyze the optical variability of AGNs with an unprecedented time resolution. Studying the asymmetry in variations could give independent constraints on the physical models for AGN variability. In this paper, we use Kepler observations of 19 sources to perform analyses on the variability asymmetry of AGNs. We apply smoothing-correction to light curves to deduct the bias to high frequency variability asymmetry, caused by long term variations which are poorly sampled due to the limited length of light curves. A parameter $\beta$ based on structure functions is introduced to quantitively describe the asymmetry and its uncertainty is measured using extensive Monte-Carlo simulations. Individual sources show no evidence of asymmetry at timescales of $1\sim20$ days and there is not a general trend toward positive or negative asymmetry over the whole sample. Stacking data of all 19 AGNs, we derive averaged $\overline{\beta}$ of 0.00$\pm$0.03 and -0.02$\pm$0.04 over timescales of 1$\sim$5 days and 5$\sim$20 days, respectively, statistically consistent with zero. Quasars and Seyfert galaxies show similar asymmetry parameters. Our results indicate that short term optical variations in AGNs are highly symmetric.
\end{abstract}



\section{Introduction}
	Aperiodic optical/ultraviolet variability is a significant property of Active Galactic Nuclei (AGNs), but its physical origin is still unclear. Variability asymmetry describes whether the light curve favors a shape of rapid rise and gradual decay, i.e., positive asymmetry, or a shape of gradual rise and rapid decay, i.e., negative asymmetry. There are only a few observational/theoretical works in literature studying the variability asymmetry of AGNs. \citet{kawaguchi98} introduced a structure function approach to estimate the variability asymmetry of AGN light curves. They adopted two structure functions, i.e., $SF_{ic}(\tau)$ and $SF_{dc}(\tau)$, which only include pair epochs with increasing and decreasing flux, respectively. Possible physical models which could produce asymmetry in variations were also discussed in this work. Through Monte Carlo simulations, \citet{kawaguchi98} showed that the disk instability model produces $SF_{ic}(\tau)<SF_{dc}(\tau)$ or negative asymmetry, while the starburst model, which attributes the optical variations to random superposition of supernovae in the nuclear starburst region, yields a contrary asymmetry, i.e., $SF_{ic}(\tau)>SF_{dc}(\tau)$. Shortly thereafter, \citet{hawkins02} added that the micro-lensing model predicts $SF_{ic}(\tau)=SF_{dc}(\tau)$ or no asymmetry statistically. 

	Observationally, \citet{hawkins02} tested the aforementioned three models using long term optical light curves of 401 quasars and 45 Seyfert galaxies. They found that Seyfert galaxy NGC 5548 appears negative asymmetric variations on timescales of $10\sim70$ days, while for quasars the variations are symmetric on timescales of a year and longer. \citet{deVries05} studied the long term variations of a large sample of 41391 quasars with SDSS and historic photometry, and showed that on timescales of years the quasar variations behave positive asymmetry. \citet{bauer09} analyzed the optical variability of nearly 23000 quasars in the Palomar-QUEST Survey and found no evidence of any asymmetry in variability over $\sim10$ days to several years. \citet{voevodkin10} computed the structure functions for 7562 quasars from SDSS Stripe 82 and detected a significant negative asymmetry on timescales longer than 300 days. All the analyses above were based on structure function method introduced by \citet{kawaguchi98}. Meanwhile, \citet{giveon99} made use of a different method of calculating the difference between medians of brightening phases and fading phases in the light curves of 42 PG quasars and reported a negative asymmetry in variations. To summarize, negative asymmetry is favored by more works, but inconsistencies exist among the few observational studies in literature.

	The Kepler space telescope was designed to search for exoplanets (\citealt{borucki10}) and it could produce nearly continuous optical light curves for the targets within its field of view, including AGNs. There are a couple of works in literature reporting the high frequency variability analyses of Kepler AGNs. \citet{mushot11} calculated the power spectral density (PSD) functions for four Seyfert 1 galaxies in the Kepler field and obtained best-fit PSD power-law slopes of $-2.6\sim-3.3$, considerably steeper than those of quasars at timescales of months to years which could be described as damped random walk process \citep[e.g.,][]{Kelly2009}. Complemented with ground based observations, \citet{carini12} presented a further analysis of the Kepler light curve of Zw 229-15. \citet{wehrle13} and \citet{revalski13} reported four radio-loud Kepler AGNs and calculated the PSD functions using light curves stitched by a normalization method. \citet{edelson13} reported the variability analyses of a BL Lac object, i.e., W2R 1926+42, in the Kepler field. 

	Meanwhile, the Kepler data provide the first ever opportunity to study the short term variability asymmetry of AGNs, which is the aim of this work. We describe the adopted Kepler AGN sample and light curve stitching process in Section \ref{sec:data}. The methods of asymmetry analysis and smoothing-correction are introduced in Section \ref{sec:methods}. In Section \ref{sec:results} we report our major results. Discussion is given in Section \ref{sec:discussion} and conclusions in Section \ref{sec:conclusion}. 

\section{Data Reduction}\label{sec:data}
\subsection{Kepler AGN Source}\label{subsec:source}
	There are only a few cataloged AGNs in the Kepler's field of view. The 19 AGNs we used in this work and listed in Table \ref{tab:all-beta-source} are collected from literature (\citealt{mushot11}; \citealt{edelson12}; \citealt{wehrle13}). Following \citet{hawkins02}, we divide the AGNs into two subsamples, including 9 quasars with $M_J<-23.6$ and 9 Seyfert galaxies with $M_J>-23.6$. The rest source (W2R 1926+42) in the table is a BL Lac object.

	We adopt the SAP FLUX in the Kepler light curve files, since PDCSAP FLUX (calibrated for systematic effects, e.g., pointing and focus changes) may have masked out the intrinsic AGN variations \citep{carini12}. 

\subsection{Stitching Light Curves} \label{subsec:stitching}
	As a result of the Differential Velocity Aberration effect of Kepler telescope, the target flux is continuously redistributed among neighboring pixels, appearing as artificial long term variations in electron counts within the fixed optimal aperture and discontinuous light curves between adjacent quarters (\citealt{kplr-instrument-handbook}; \citealt{kinemuchi12}). \citet{carini12} rescaled the Kepler light curves with coordinated ground based observations to connect different quarters, however, it is not a general approach. \citet{kinemuchi12} gave another method, which increases the number of pixels in target mask for photometry with the PyKE tasks $kepmask$ and $kepextract$. With reset target mask, a new light curve can be extracted from the counterpart Target Pixel File. The later method could reduce target flux losses out of the aperture, but it might also get contamination from nearby sources. 

	\begin{figure}[!ht]
		\centering
		\includegraphics[width=\columnwidth]{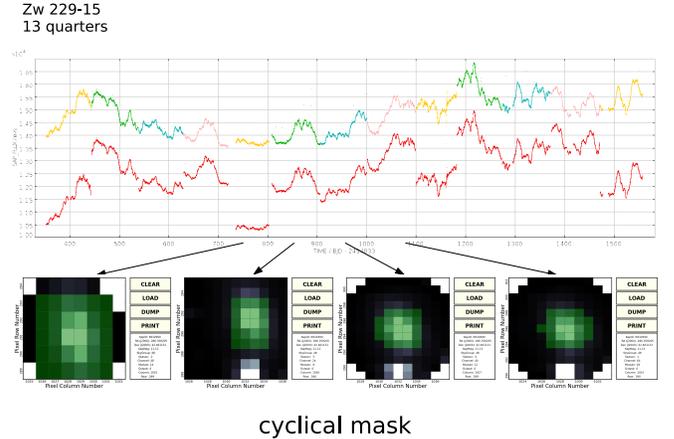}
		\caption{The original and stitched light curves with bin size of 30 minutes (hereafter the same) of Zw 229-15. Vertical axis denotes the flux (count rate). The original data is shown in red, while the re-extracted light curves are shown in yellow, green, blue, and pink. Four reset target masks are adopted (plotted with different colors in the light curves) and duplicately used for four quarters in each year. The four masks are shown at the bottom for demonstration. \label{fig:reextracted}}
	\end{figure}

	Following \citet{kinemuchi12}, we perform the pixels re-extraction process as illustrated in Figure \ref{fig:reextracted}. Since not all the sources have enough observational quarters and clear surrounding and, during the process, enough `halo' pixels surrounding the target are needed, only four sources, i.e., Zw 229-15, W2 1925+50, W2R 1904+37, and CGRaBS J1918+4937, can be stitched. The results are shown in Section \ref{subsec:stitched-results}.

\section{Analysis Methods}\label{sec:methods}
\subsection{Structure Function and Asymmetry Parameter}\label{subsec:beta-definiton} 
	The general definition of structure function and its properties were given by \citet{simonetti85}. The first-order structure function $SF(\tau)$ is defined as
	\begin{equation}
		SF^2(\tau)=\frac{1}{N(\tau)}\sum_t{[f(t+\tau)-f(\tau)]^2}, \label{equ:sf}
	\end{equation}
where $N(\tau)$ is the number of data pairs for a certain time lag $\tau$, and $f$ the observed source flux\footnote{Computing the structure function and then the asymmetry parameter using magnitude does not alter the results presented in this work.}. To quantify variability asymmetry, an asymmetry parameter $\beta(\tau)$ is defined as \citep[see][]{kawaguchi98}
	\begin{equation}
		\beta(\tau)=\frac{SF_{ic}(\tau)-SF_{dc}(\tau)}{SF_{tot}(\tau)}, \\\label{equ:beta}
	\end{equation}
where the suffix $ic$ and $dc$ denote data pairs with increasing and decreasing flux, respectively, and $tot$ for the total data pairs. $\beta(\tau)$ quantifies the normalized difference between the increasing and decreasing variability. A positive $\beta(\tau)$ represents that the light curve favors a shape of rapid rise and gradual decay, i.e., positive asymmetry, while a negative $\beta(\tau)$ characterizes the opposite situation. 

\subsection{Smoothing-Correcttion}\label{subsec:smooth}
\begin{figure}[!t]
		\includegraphics[width=\columnwidth]{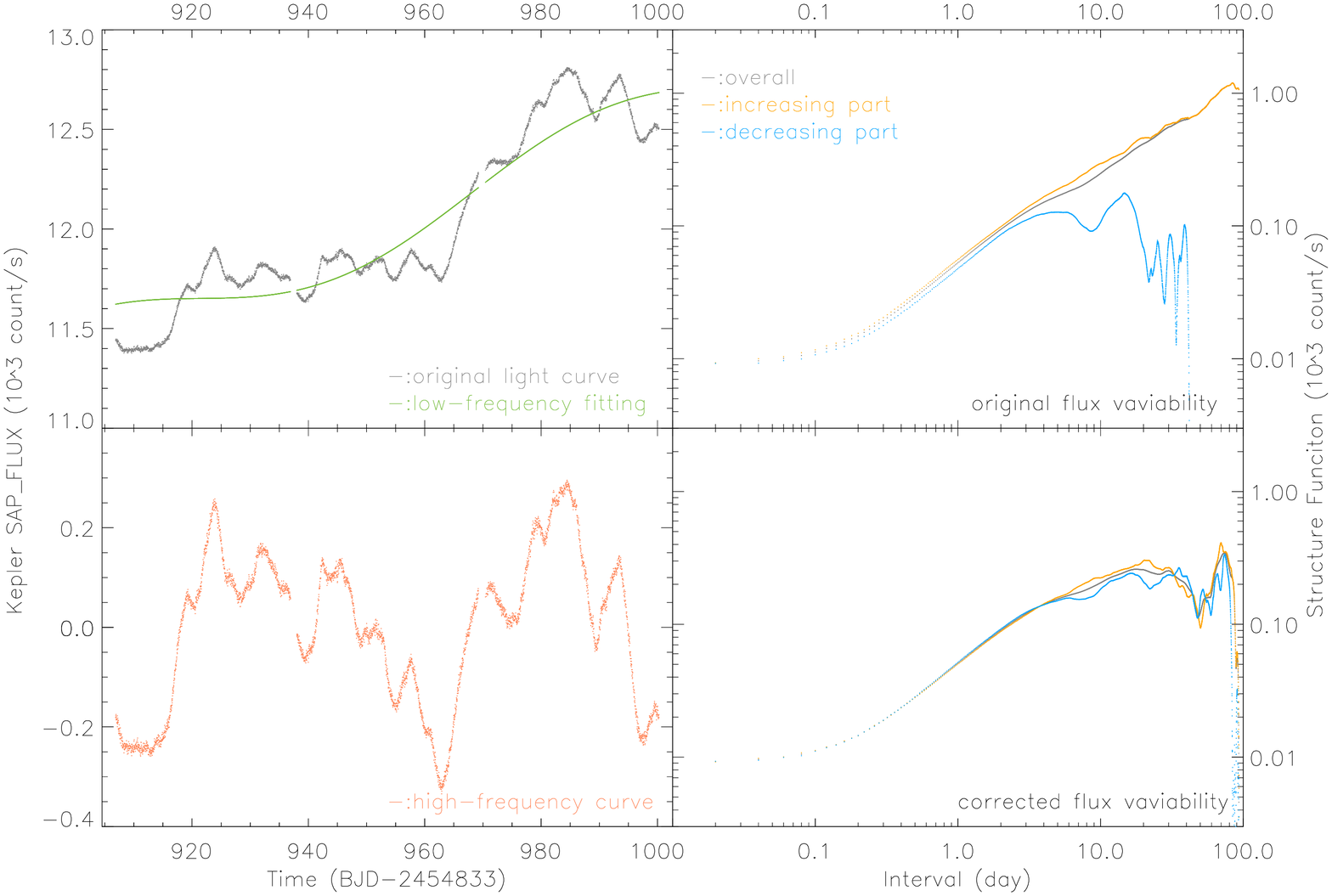}
		\figcaption{An example shows the smoothing-correction applied to the tenth quarter of Zw 229-15. The left two panels show light curves with the vertical axis denoting the flux (count rate), where the original (grey) and smoothing-corrected high-frequency (red) light curves are shown in the upper and lower panels, respectively. The low-frequency curve (green) in the upper-left panel is the difference between the original and the high frequency curves. The right two panels show the corresponding structure functions. $SF_{ic}(\tau)$ and $SF_{dc}(\tau)$ are shown in orange and blue, separately, and the total $SF_{tot}(\tau)$ in grey. Strong positive asymmetry, i.e., $SF_{ic}(\tau)$ $>$ $SF_{dc}(\tau)$, is visible at timescales above a couple of days in upper-right panel, due to the long term increasing trend seen in the original light curve. Weak positive asymmetry is also visible at shorter timescales in the original structure function (upper-right panel), but disappears after smoothing-correction (lower-right panel). \label{fig:smooth}}
	\end{figure}

	Since the light curves have limited duration, the long term variations are poorly sampled and will yield unphysical and large scatter in the measurement of the asymmetry parameter. In other words, for individual light curves, asymmetry analyses could only be performed to short term variations (the ratio of the duration of the light curve to the concerned timescale of variations $\gg$ 1), which have been well sampled.  Furthermore, the long term trend in the light curves could also introduce biases into the short term asymmetry analyses. 
	Such effect can be reduced using smoothing-correction method, in which the corrected light curve is produced by the inverse Fourier transform of the power spectrum with the low frequency part set to zero. The cutoff frequency in the power spectrum is set to 2$\times$10$^{-7}$ Hz, which corresponds to about 60 days and is shorter than the duration of light curves for each quarter of all sources. An example is illustrated in Figure \ref{fig:smooth} and Section \ref{sec:results} presents the results of asymmetry analysis for both the original and smoothing-corrected light curves.
	
\subsection{Uncertainty Estimation}\label{subsec:simulate}
	Estimating the uncertainties in $SF(\tau)$ and $\beta(\tau)$ is not straightforward \citep[see][]{emma10}. The $[f(t+\tau)-f(\tau)]$ series (in Equation \ref{equ:sf}) of a single light curve are not mutually independent, thus the uncertainties in $SF(\tau)$ and $\beta(\tau)$ at given $\tau$ would be significantly underestimated with the standard deviation definition. Furthermore, the values of $SF(\tau)$ and $\beta(\tau)$ at different timescales $\tau$ are not independent either. Therefore, following \cite{emma10}, we estimate the uncertainties through extensive Monte-Carlos simulations. 
	
	We adopt the algorithm of \citet{timmer95} to generate artificial light curves for given power spectrum. In order to make the simulated light curves possessing similar power spectrum shape and noise level with the observed ones, we use the original periodogram, calculated from the original observed (and not the smoothing-corrected) light curves, as the input spectrum, instead of using power spectrum with a specific fixed power-law slope. The observed light curve is end-matched before periodogram calculation to reduce contamination caused by the mismatch between beginning and end points \citep[e.g.,][]{mushot11,wehrle13,edlson14}
	\footnote{Without end-matching the high frequency components of the power spectrum will be spuriously enhanced, then the spectrum slope will be flatter than reality \citep{fougere85,mart12}. The simulated light curves, based on such input power spectrum, will be significantly biased against the observed one. Note that in this work the end-matching was only adopted to calculate the power spectrum. No end-matching is performed for calculations of structure function and asymmetry parameter.}. 
	To simulate the effect of red noise leak to the short light curves, the artificial light curves should be much longer than the observed ones, by extending the input power spectrum to lower frequencies \citep[e.g.,][]{uttley02,vaughan03,emma10}. We fit the low frequency part of the original power spectrum to measure the low frequency extending slope. If the best-fitted slope is steeper than -2, we adopt a fixed value of -2, as seen in the observed (lower frequency comparing with Kepler) power spectrum of quasars \citep[e.g.,][]{Kelly2009}
	\footnote{Note that adopting steeper lower frequency power spectrum slope would yield slightly larger scatter in the derived $SF(\tau)$ and $\beta(\tau)$, due to stronger red noise leak. However, the asymmetric scatter of smoothing-correction data, which we mostly care about, will not be influenced by the enhanced low frequency leak.}. 
	
	For each observed Kepler light curve, we generate a single, 3000 times longer light curve, which is then split into 3000 segments. We then randomly select 1000 segments of them to calculate the corresponding $SF(\tau)$ and $\beta(\tau)$, following exactly the same procedures as we applied on the observed light curves, and take their scatters as the uncertainties of the observed $SF(\tau)$ and $\beta(\tau)$, respectively. Note that during the simulations, it has been assumed that the variations are intrinsically symmetric, thus the output mean of simulated $\beta(\tau)$ equals zero, and its scatter represents the uncertainty of the observed $\beta(\tau)$ if not severely different from zero.

\section{Variability Asymmetry Results}\label{sec:results}
\subsection{Results of Quarterly-Stitched Light Curves}\label{subsec:stitched-results}
\begin{figure}[!t]
		\includegraphics[width=\columnwidth]{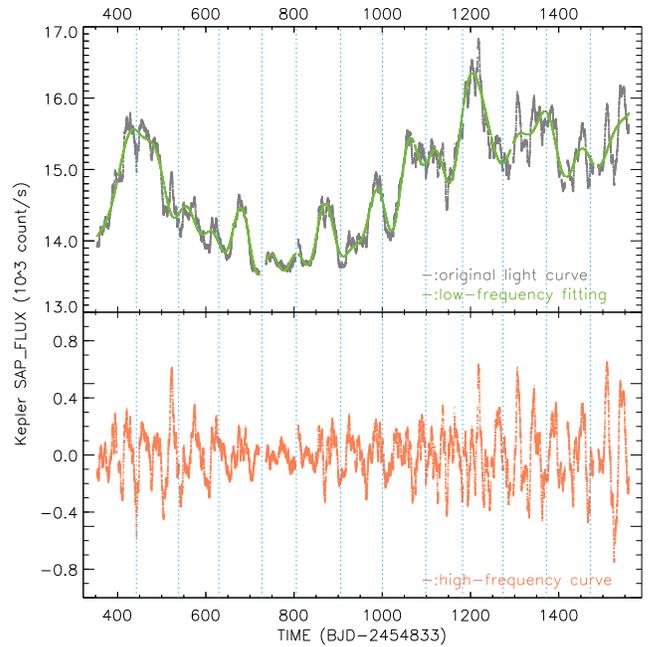}
		\figcaption{The stitched light curves of Zw 229-15. The vertical axis denotes the flux (count rate). The original (grey) and smoothing-corrected high-frequency (red) light curves are shown in the upper and lower panels, respectively. The low-frequency curve (green) in the upper panel is the difference between the original and the high frequency curves. Blue dotted lines represent boundaries of intervals of different quarters. \label{fig:zw229-15-lc-eps}}
	\end{figure}

	In this section, we present the results from asymmetry analysis of the stitched light curves, constructed for the four sources listed in Section \ref{subsec:stitching} (i.e., Zw 229-15, W2 1925+50, W2R 1904+37, and CGRaBS J1918+4937). The stitched light curve of Zw 229-15, spanning $\sim3.3$ years with a time resolution of 30 minutes, is presented in Figure \ref{fig:zw229-15-lc-eps}. Zw 229-15 is a narrow-line Seyfert 1 galaxy, observed by Kepler during 13 quarters (Q4$\sim$Q16, the most among Kepler AGNs). 

\begin{figure}[!t]
		\includegraphics[width=\columnwidth]{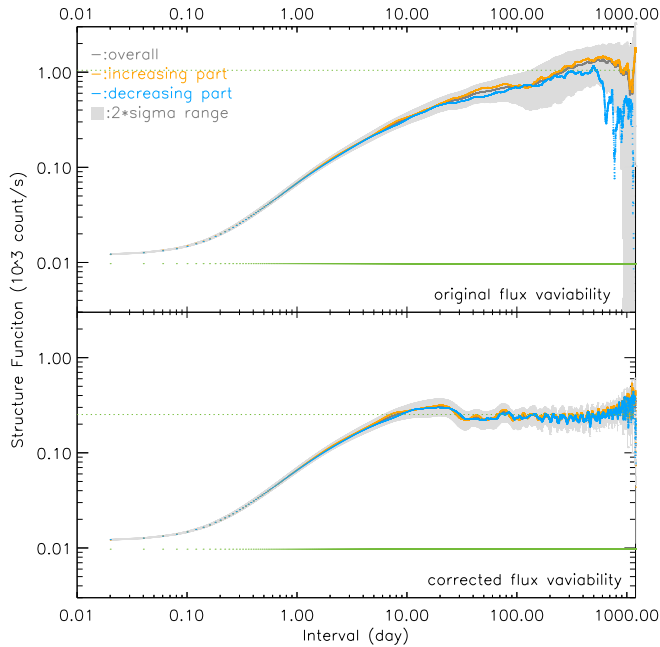}
		\figcaption{$SF(\tau)$ of stitched light curves of Zw 229-15. $SF_{ic}(\tau)$ and $SF_{dc}(\tau)$ are shown in orange and blue, separately, and the total $SF_{tot}(\tau)$ in grey; shadow areas represent $2\sigma$ error range of $SF_{tot}(\tau)$. The top panel is for the original data and the bottom for the smoothing-corrected data. \label{fig:zw229-15-sf-eps}}
	\end{figure}

	Figure \ref{fig:zw229-15-sf-eps} shows structure functions of both the original and smoothing-corrected light curves of Zw 229-15. The original $SF(\tau)$ shows a rise towards longer timescales with a gradually decreasing slope. At timescales above 200 days, $SF_{ic}(\tau)$ is significantly larger than $SF_{dc}(\tau)$, consistent with the general increasing trend in the original light curve. At very short timescales, the $SF(\tau)$ reaches a plateau due to the photometric uncertainty of the light curve, with value of $\sqrt{2}\sigma_m$, where $\sigma_m$ is the photometric uncertainty. On the long term end, $SF(\tau)$ converges to $\sqrt{2}\sigma_l$, where $\sigma_l$ is the standard deviation of whole light curve. As a result of the smoothing-correction, the corrected $SF(\tau)$ becomes remarkably flat at long timescales, and equals $\sqrt{2}\sigma_l$ at timescales longer than $\sim20$ days, which appears as an `ideal' $SF(\tau)$ where the related light curve is long enough that the edge effects and aliasing are negligible (\citealt{hughes92}). From the smoothing-corrected $SF(\tau)$, we see no clear asymmetry in the variations.
	
\begin{figure}[!t]
		\includegraphics[width=\columnwidth]{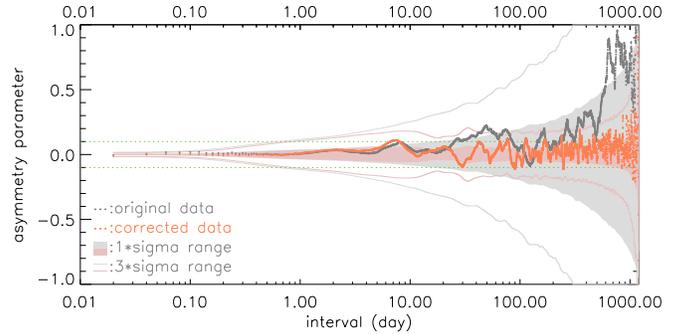}
		\figcaption{$\beta(\tau)$ of stitched light curves of Zw 229-15. The one from original data is shown in grey and that after smoothing-correction in red. The $1\sigma$ and $3\sigma$ uncertainty ranges of $\beta(\tau)$, measured from simulated light curves assuming the variations are intrinsically symmetric, i.e.,  with mean $\beta(\tau)$ = 0, are shown in filled areas and narrow lines with light-color, respectively. The dotted green line marks $\beta(\tau)=\pm0.1$. \label{fig:zw229-15-beta-eps}}
	\end{figure}

\begin{figure}[!t]
		\centering
		\includegraphics[trim=0cm 1.39cm 0cm 0cm,clip,width=\columnwidth]{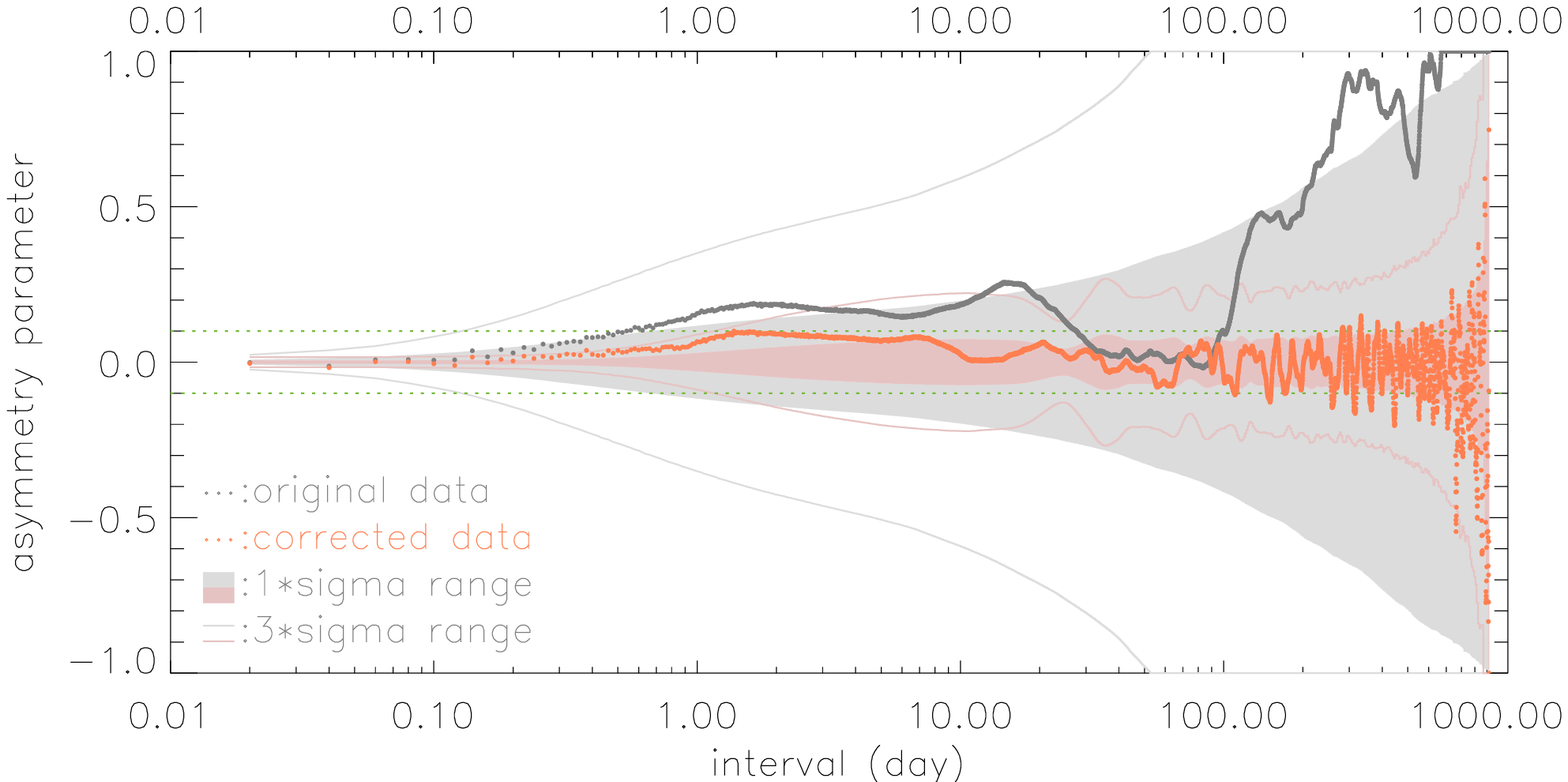}
		\includegraphics[trim=0cm 1.39cm 0cm 0.59cm,clip,width=\columnwidth]{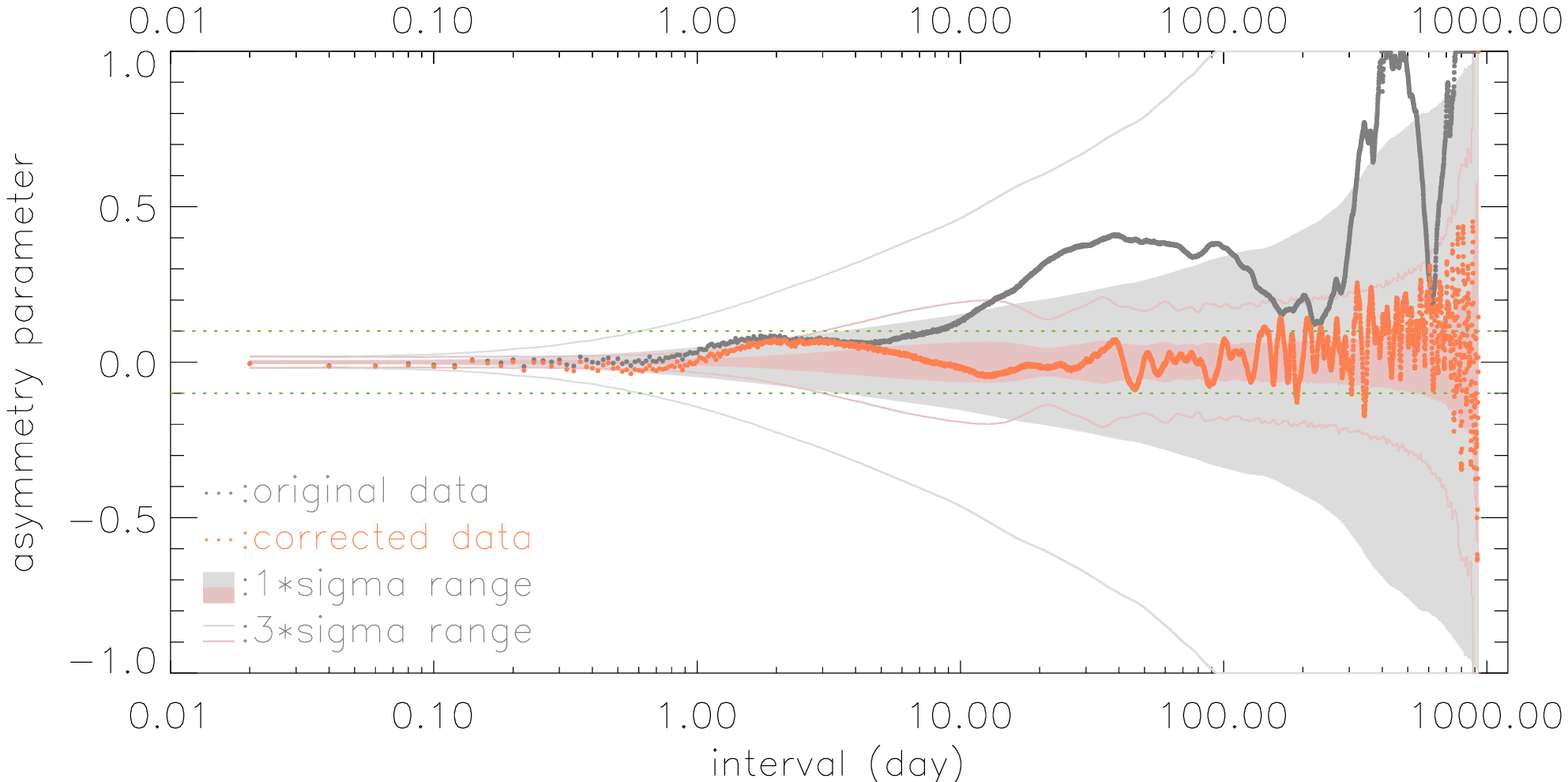}
		\includegraphics[trim=0cm 0cm 0cm 0.59cm,clip,width=\columnwidth]{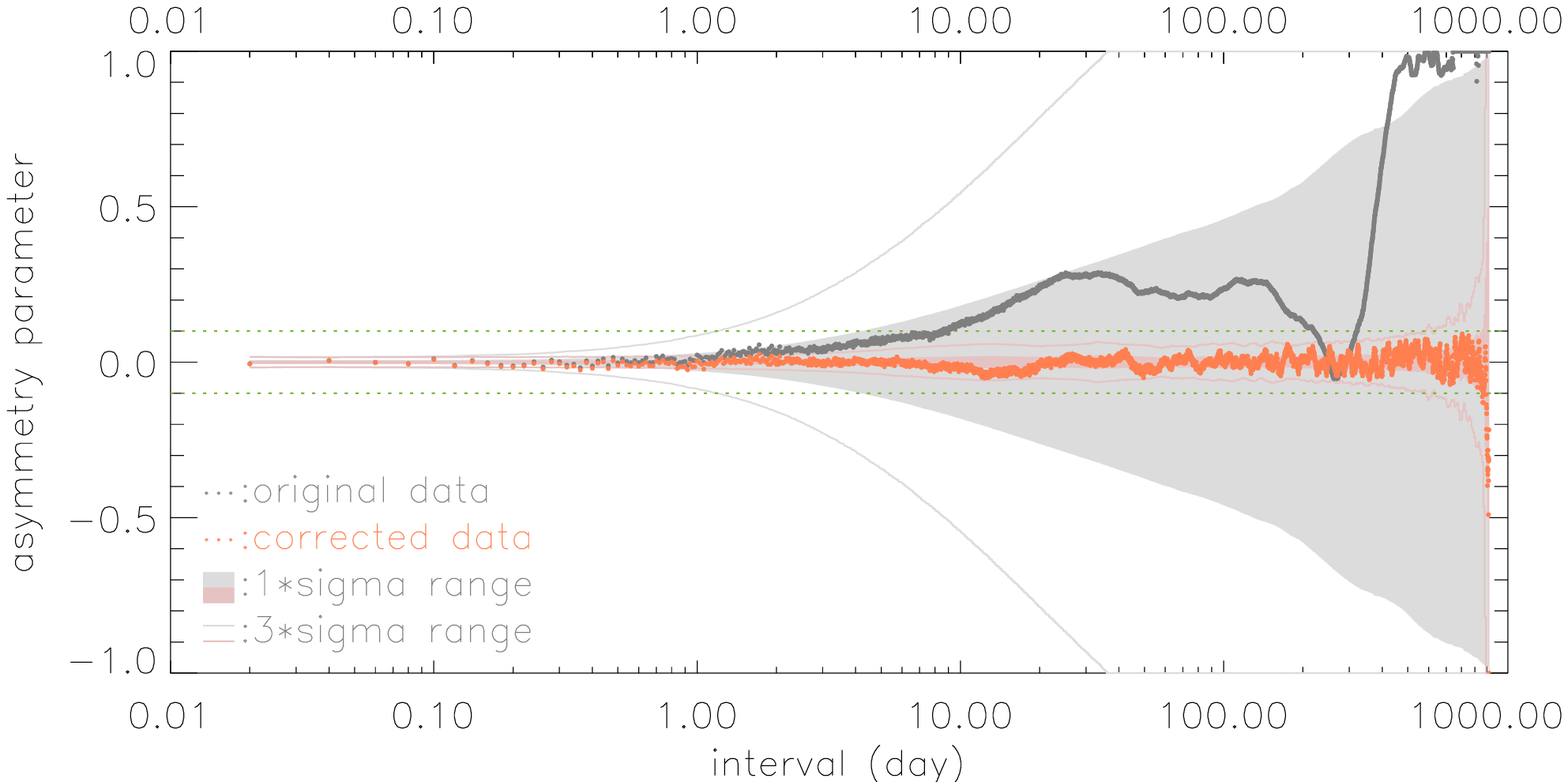}
		\figcaption{$\beta(\tau)$ of stitched light curves of W2 1925+50(top), W2 1925+50(middle), and CGRaBS J1918+4937(bottom). The figure legends are the same as Figure \ref{fig:zw229-15-beta-eps}. \label{fig:other3-beta-eps}}
	\end{figure}

	We plot the $\beta(\tau)$ curves for both the light curves in Figure \ref{fig:zw229-15-beta-eps}. The original $\beta(\tau)$ curve has a positive excess at long timescales, while the smoothing-corrected $\beta(\tau)$ curve appears approximately with $|\beta(\tau)|<0.1$ at the whole range. The $\beta(\tau)$ curves of the other three stitched sources are shown in Figure \ref{fig:other3-beta-eps}, all of which appear similar to Zw 229-15, and we find no evidence of asymmetry after smoothing-correction. 

\subsection{Averaged Asymmetry Parameter of the Sample}
	We have analyzed 19 Kepler AGNs, each of which has $3\sim13$ observational quarters. For each AGN, we derive $\beta(\tau)$ curve for each of the observational quarter, and average them to get $\overline{\beta}(\tau)$ for each source. There is an alternative process, in which we average structure functions of different quarters first and then derive $\overline{\beta}(\tau)$ from $\overline{SF}(\tau)$ with Equation \ref{equ:beta}. However, with the second approach, the averaged asymmetry could be dominated by the quarter(s) with larger variation amplitudes, while with the first method the asymmetry parameter is simply averaged over time. For the same reason, no weight is introduced during averaging. 

	\begin{table*}[!ht]
		\centering
		\begin{threeparttable}[b]
		\caption{The time-mean $\overline{\beta}_x$ of all AGNs.\label{tab:all-beta-source}}
		\begin{tabular*}{1.0\textwidth}{@{\extracolsep{\fill}} c r c c c c c c c c c c l }
			\hline
			\hline
			\multirow{2}{*}{Source Name} &\multirow{2}{*}{Kepler ID} &\multirow{2}{*}{RA} &\multirow{2}{*}{Dec} &\multirow{2}{*}{z} &\multirow{2}{*}{$n(q)$\tnote{d}} &\multicolumn{2}{c}{original} &\multicolumn{2}{c}{smoothing-corrected} &\multirow{2}{*}{type}\\
			& & & & & &$\overline{\beta}_{1\sim5}$ &$\overline{\beta}_{5\sim20}$ &$\overline{\beta}_{1\sim5}$ &$\overline{\beta}_{5\sim20}$ & \\ \hline
			Zw 229-15\tnote{a}  &6932990  &19 05 26.0 &+42 27 40 &0.028 &13  &\phantom{-}0.04$\pm$0.07 &\phantom{-}0.10$\pm$0.11  &\phantom{-}0.01$\pm$0.05 &-0.01$\pm$0.05  &Sy1\\
			W2 1925+50\tnote{a} &12158940 &19 25 02.2 &+50 43 14 &0.067 &11  &\phantom{-}0.04$\pm$0.10 &-0.04$\pm$0.12  &\phantom{-}0.02$\pm$0.06 &-0.02$\pm$0.06 &Sy1\\
			W2R 1858+48         &11178007 &18 58 01.1 &+48 50 23 &0.079 &8   &-0.23$\pm$0.09 &-0.24$\pm$0.12  &-0.10$\pm$0.06 &-0.01$\pm$0.07 &Sy1\\
		$\ $W2R 1904+37\tnote{a}&2694186  &19 04 58.7 &+37 55 41 &0.089 &10  &\phantom{-}0.03$\pm$0.08 &\phantom{-}0.09$\pm$0.13  &-0.04$\pm$0.04 &-0.04$\pm$0.06 &Sy1\\
			W2R 1914+42              &6595745  &19 14 15.5 &+42 04 59 &0.502 &6   &-0.03$\pm$0.08 &\phantom{-}0.06$\pm$0.15  &\phantom{-}0.00$\pm$0.04 &\phantom{-}0.09$\pm$0.07 &QSO\tnote{b} \\
			W2R 1920+38              &3337670  &19 20 47.7 &+38 26 41 &0.368 &6   &\phantom{-}0.10$\pm$0.10 &\phantom{-}0.18$\pm$0.17  &\phantom{-}0.00$\pm$0.04 &\phantom{-}0.00$\pm$0.06 &QSO\tnote{b} \\
			W2R 1931+43              &7610713  &19 31 12.5 &+43 13 27 &0.439 &7   &\phantom{-}0.00$\pm$0.12 &\phantom{-}0.00$\pm$0.18  &-0.03$\pm$0.05 &-0.04$\pm$0.07 &QSO\tnote{b,c} \\
			W2R 1910+38              &2837332  &19 10 02.5 &+38 00 09 &0.130 &6   &\phantom{-}0.04$\pm$0.10 &\phantom{-}0.08$\pm$0.16  &-0.05$\pm$0.06 &-0.04$\pm$0.08 &Sy1\tnote{b} \\
			W2R 1853+40              &5597763  &18 53 19.2 &+40 53 36 &0.625 &5   &\phantom{-}0.01$\pm$0.12 &-0.07$\pm$0.19  &\phantom{-}0.01$\pm$0.05 &-0.04$\pm$0.07 &QSO\tnote{b} \\
			W2R 1845+48              &10841941 &18 45 59.5 &+48 16 47 &0.152 &6   &\phantom{-}0.05$\pm$0.08 &\phantom{-}0.11$\pm$0.15  &\phantom{-}0.00$\pm$0.05 &-0.06$\pm$0.07 &QSO\tnote{b} \\
			W2R 1931+38              &3347632  &19 31 15.4 &+38 28 17 &0.158 &6   &\phantom{-}0.11$\pm$0.11 &\phantom{-}0.02$\pm$0.17  &\phantom{-}0.14$\pm$0.06 &\phantom{-}0.08$\pm$0.07 &Sy1\tnote{b} \\
			W2R 1926+42              &6690887  &19 26 31.0 &+42 09 59 &0.154 &6   &\phantom{-}0.00$\pm$0.04 &-0.03$\pm$0.07  &\phantom{-}0.00$\pm$0.03 &\phantom{-}0.04$\pm$0.05 &BL Lac\\
			KA 1915+41	             &5781475  &19 15 09.1 &+41 02 39 &0.220 &3   &-0.27$\pm$0.20 &-0.37$\pm$0.28  &\phantom{-}0.00$\pm$0.10 &-0.13$\pm$0.12 &QSO\tnote{b} \\
			KA 1922+45               &9215110  &19 22 11.2 &+45 38 06 &0.115 &7   &-0.07$\pm$0.06 &\phantom{-}0.09$\pm$0.11  &-0.05$\pm$0.04 &\phantom{-}0.01$\pm$0.06 &Sy1.9\\
			1RXS J192949.7+462231    &9650715  &19 29 50.5 &+46 22 24 &0.127 &4   &-0.10$\pm$0.16 &-0.14$\pm$0.21  &\phantom{-}0.09$\pm$0.09 &\phantom{-}0.04$\pm$0.09 &Sy1\\
			MG4 J192325+4754         &10663134 &19 23 27.2 &+47 54 17 &1.520 &11  &\phantom{-}0.01$\pm$0.04 &\phantom{-}0.06$\pm$0.09  &\phantom{-}0.00$\pm$0.02 &\phantom{-}0.02$\pm$0.04 &QSO\\
			MG4 J190945+4833         &11021406 &19 09 46.5 &+48 34 32 &0.513 &8   &-0.15$\pm$0.05 &-0.32$\pm$0.11  &-0.01$\pm$0.01 &-0.06$\pm$0.03 &QSO\\
			CGRaBS J1918+4937\tnote{a} &11606854 &19 18 45.6 &+49 37 55 &0.926 &11  &\phantom{-}0.04$\pm$0.06 &\phantom{-}0.14$\pm$0.11  &-0.02$\pm$0.03 &-0.10$\pm$0.05 &QSO\\
			{}[HB89] 1924+507        &12208602 &19 26 06.3 &+50 52 57 &1.098 &11  &-0.15$\pm$0.03 &-0.35$\pm$0.08  &\phantom{-}0.01$\pm$0.01 &\phantom{-}0.00$\pm$0.03 &Sy1.5\\
			\hline
		\end{tabular*}
		\begin{tablenotes}
			\item [a] The four AGNs have stitched light curves, which have been analyzed in Section \ref{subsec:stitched-results}.
			\item [b] The eight AGNs are classified as QSOs or Seyfert galaxies according to their $M_J$.
			\item [c] \citet{edelson13} classified W2R 1931+43 as a Seyfert 1 galaxy.
			\item [d] The column shows number of observational quarters for each Kepler AGN (the same in Table \ref{tab:mix-beta}).
		\end{tablenotes}
		\end{threeparttable}
	\end{table*}

	The mean $\overline{\beta}$ over timescales of $1\sim5$ days and $5\sim20$ days are also derived and listed in Table \ref{tab:all-beta-source}. The uncertainties of $\overline{\beta}_{1\sim5}$ and $\overline{\beta}_{5\sim20}$ for individual sources are also calculated through simulations. Again, after smoothing-correction, most of the sources show very small asymmetry parameter (both positive and negative values are seen with $|\overline{\beta}|<0.1$), consistent with zero within the statistical uncertainties. For the whole sample there is not a general trend towards a positive or negative asymmetry. This indicates there is no or at most very weak variability asymmetry in Kepler AGNs.

\begin{figure}[!ht]
		\includegraphics[width=\columnwidth]{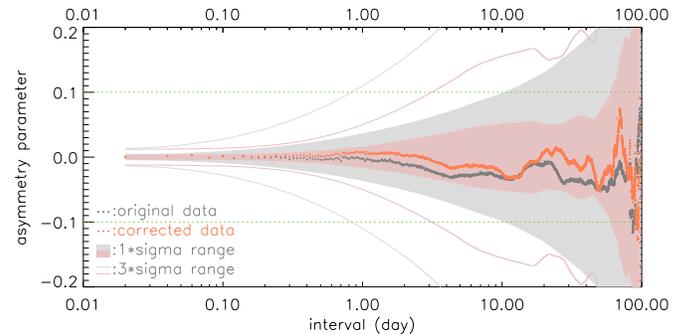}
		\figcaption{The average $\overline{\beta}(\tau)$ of the whole sample including all sources. The figure legends are the same as Figure \ref{fig:zw229-15-beta-eps}. 
		\label{fig:mix-beta-eps}
		}
	\end{figure}

	To derive better constraint on the asymmetry parameter, we average $\beta(\tau)$ from all quarters of all sources. The uncertainties in $\overline{\beta}(\tau)$ are also derived from simulations\footnote{Since there is a large number (145) of Kepler quarters (real data), we can measure the intrinsic scatter of $\beta(\tau)$ from different quarters and derive the uncertainties in the averaged $\overline{\beta}(\tau)$. The uncertainties in $\overline{\beta}(\tau)$ derived with this approach are slightly smaller than those from simulations, by a factor of 1.5 -- 2.0. In this work, we adopt the more conservative measurements from simulations. Nevertheless, our conclusions are not affected by the selection.}. The co-added $\overline{\beta}(\tau)$ averaged over all sources and its uncertainty are plotted in Figure \ref{fig:mix-beta-eps}. There is little difference between the original and smoothing-corrected $\overline{\beta}(\tau)$ curves in Figure \ref{fig:mix-beta-eps}, which implies that the bias due to long term variations with limited duration of light curves has been extensively reduced after averaging a large number (145) of quarters. However, we note the smoothing-corrected co-added $\overline{\beta}(\tau)$ has considerably smaller scatter (Figure \ref{fig:mix-beta-eps}), as the scatter due to long term variations has been reduced.

	The $\overline{\beta}(\tau)$ averaged over timescales of $1\sim5$ days and $5\sim20$ days of the whole sample and two subsamples (quasars and Seyfert galaxies) are listed in Table \ref{tab:mix-beta}. We see that the values of the co-added $\overline{\beta}(\tau)$ on timescales of $1\sim20$ days are all consistent with zero within the small statistical uncertainties (again derived with simulations), suggesting the variations are highly symmetric.

	\begin{table*}[!ht]
		\caption{The time-mean $\overline{\beta}_x$ of mixed samples.\label{tab:mix-beta}}
		\centering
		\begin{tabular*}{1.0\textwidth}{@{\extracolsep{\fill}} c r c c c c }
			\hline
			\hline
			\multirow{2}{*}{Sample} &\multirow{2}{*}{$n(q)$} &\multicolumn{2}{c}{original} &\multicolumn{2}{c}{smoothing-corrected}\\
			& &$\overline{\beta}_{1\sim5}$ &$\overline{\beta}_{5\sim20}$ &$\overline{\beta}_{1\sim5}$ &$\overline{\beta}_{5\sim20}$\\ \hline
			Quasars           &63   &-0.01$\pm$0.06 &\phantom{-}0.00$\pm$0.12  &-0.01$\pm$0.03 &-0.03$\pm$0.05 \\
			Seyfert Galaxies  &76   &-0.03$\pm$0.06 &-0.04$\pm$0.10  &\phantom{-}0.00$\pm$0.04 &-0.01$\pm$0.05 \\
			All Sources       &145  &-0.02$\pm$0.06 &-0.02$\pm$0.10  &\phantom{-}0.00$\pm$0.03 &-0.02$\pm$0.04 \\
			\hline
		\end{tabular*}
	\end{table*}

\section{Discussion}\label{sec:discussion}
	Using all Kepler light curves of 19 AGNs, we find the variability asymmetry at timescales of $1\sim20$ days is rather weak. The averaged $\overline{\beta}$ is 0.00$\pm$0.03 and -0.02$\pm$0.04 over timescales of 1$\sim$5 days and 5$\sim$20 days, respectively, statistically consistent with zero.

	It's convenient to interpret the asymmetry parameter in terms of shot-noise model, in which  the variations in AGNs are attributed to the stochastic superposition of independent discrete flares \citep[e.g.,][]{negoro95}. Such model provides a mathematical framework for a series of physical models \citep{cid00,favre05}, including starburst and micro-lensing \citep{kawaguchi98,hawkins02}. Generally, a positive asymmetry parameter $\beta$ indicates the flare rises rapidly and decays gradually, such as expected in the starburst model, which attributes the variations in AGNs to random superposition of supernovae in the nuclear starburst region. Clearly, our detection of highly symmetric variations in Kepler AGNs does not favor the starburst model. Micro-lensing model does predict symmetric variations \citep{hawkins02}, however, for our low redshift sample, the probability of micro-lensing would be insignificant \citep{hawkins02}. Actually, independent to variability asymmetry studies, these two models (starburst and micro-lensing) are disfavored by observations which show the variations of emission lines in AGNs closely correlate with but lag continuum variations \citep[e.g.,][]{Peterson2004}.

	In the disk instability (hereafter DI) model of \citet{kawaguchi98}, the optical variability is ascribed to instabilities of accretion disk as matter flows, and the asymmetry is due to single large scale avalanches. By considering a disk atmosphere emitting a power-law X-ray spectrum that fluctuating in time and assuming optical variations simply follows X-ray with little delay. \citet{kawaguchi98} made Monte-Carlo simulations adopting the cellular-automaton model of \citet{Mineshige1994} and treated the atmosphere as advection-dominated accretion flow. The simulated light curve shows an negative asymmetry with slow rise and rapid decline on timescales of one to several hundred days. As simulated by \citet{kawaguchi98}, in DI model, $\beta(\tau)=-0.1$ corresponds approximately to the ratio of diffusion mass to inflow mass of $0.1\sim0.5$ (cf. their Fig. 7), with the ratio of outer to inner disk radii fixed to be 20. Our strong constraints to the asymmetry parameter from the co-added sample therefore request an even higher ratio of diffusion mass to inflow mass based on this model.

	Theoretical calculations on the variability asymmetry parameter for more specific physical models are required to make comparison with observations. Nevertheless, the result of this work indicates the variations in AGNs rise and decay highly symmetrically. If we attribute AGN variations to perturbations in the accretion disk,  such perturbations need also behave symmetrically in both directions of time.
	
\section{Conclusions}\label{sec:conclusion}
	In this paper we use the high quality light curves from Kepler space telescope to analyze the variability asymmetry of 19 AGNs. An asymmetry parameter $\beta(\tau)$ is introduced for quantitive description. We perform extensive Monte-Carlo simulations to derive the statistical uncertainties in $\beta(\tau)$, which can not be obtained through the standard error analyses approach. After correction of observational bias due to long term trend in the light curves, we find no evidence of asymmetry in individual sources at timescales below 20 days. For the whole sample there is not a general trend towards a positive or negative asymmetry. Co-adding data for all 19 AGNs, we derive an averaged $\overline{\beta}$ of 0.00$\pm$0.03 and -0.02$\pm$0.04 over timescales of 1$\sim$5 days and 5$\sim$20 days, respectively, statistically consistent with zero. Quasars and Seyfert galaxies tend to show similar asymmetry parameter at the observed timescales. The constraint on longer timescale variability asymmetry is weaker as it requires much larger samples or much longer light curves which could better sample the long term variations. The fact that the short term optical variations in quasars and Seyfert galaxies are highly symmetric could put independent constraints on physical models of AGN variations.
	
\section*{Acknowledgement}
	We acknowledge the anonymous referee for valuable comments and constructive suggestions. This work is supported by Chinese NSF (grant No.11233002 $\&$ 11421303) and National Basic Research Program of China (973 program, grant No. 2015CB857005). J.X.W. acknowledges support from Chinese Top-notch Young Talents Program and the Strategic Priority Research Program ``The Emergence of Cosmological Structures" of the Chinese Academy of Sciences (grant No.XDB09000000). We gratefully thank Prof.Wei-Min Gu for discussion on accretion disk theories. We thank Zhen-Yi Cai for careful readings of the manuscript, the Mikulski Archive for Space Telescopes (MAST), and the Kepler team groups, especially Karen Levay, for data access. We use PyKE for stitching light curves, which is an open source suite of python software tools developed by the NASA Kepler Guest Observer Office, to reduce and analyze Kepler light curves, TPFs, and FFIs. We also thank the NASA/IPAC Extragalactic Database (NED), which is operated by the Jet Propulsion Laboratory, California Institute of Technology, under contract with the National Aeronautics and Space Administration, and the SIMBAD database, operated at CDS, Strasbourg, France, for AGN identifications.

\end{document}